\documentclass[
    ,final            % use final for the camera ready runs
  ]
  {aipproc}

\layoutstyle{8x11double}

\begin{document}

\title{Structural and Nucleosynthetic Evolution of Metal-poor 
       \& Metal-free Low and Intermediate Mass Stars}

\classification{ % http://www.aip.org/pacs/pacs08/pacs08-toc.html
                \texttt{97.10.Cv, 97.20.Wt} 
               }

\keywords{Population III, Stellar evolution, Nucleosynthesis, Metal poor halo stars}

\author{Simon W. Campbell}
       {
        address={Academia Sinica Institute of Astronomy and Astrophysics,
                 P.O. Box 23-141, Taipei 10617, Taiwan},
        email={simcam@asiaa.sinica.edu.tw}
       }

\author{John C. Lattanzio}
       {
        address={Centre for Stellar and Planetary Astrophysics,
                 School of Mathematical Sciences,
                 Monash University,
                 Melbourne, Australia 3800},
        email={john.lattanzio@sci.monash.edu.au}
}
%
%
%%%%%%
\begin{abstract}
%%%%%%
We report on an investigation into stellar evolution and nucleosynthesis
in the low and extremely low metallicity regime, including models
of stars with a pure Big Bang composition (i.e. $Z=0$). The metallicity
range of the extremely metal-poor (EMP) models we have calculated
is $-6.5<\textrm{[Fe/H]}<-3.0$, whilst our models are in
the mass range $0.85<\textrm{M}<3.0\textrm{M}_{\odot}$. Many of the
EMP and $Z=0$ models experience violent evolutionary
episodes not seen at higher metallicities. We refer to these events
as `Dual Flashes' since they are characterised by peaks in the hydrogen
and helium burning luminosities occurring at roughly the same time.
Some of the material processed by these events is later dredged up
by the convective envelope, causing significant surface pollution.
These events have been reported by previous studies, so our results
reaffirm their occurrence -- at least in 1D stellar models. The novelty
of this study is that we have calculated the entire evolution of the
$Z=0$ and EMP models, from the ZAMS to the end of the TPAGB, including
detailed nucleosynthesis. We have also calculated the nucleosynthetic
yields, which will soon be available in electronic format. Although
subject to many uncertainties these are, as far as we are aware, the
only yields available in this mass and metallicity range. In this 
paper we briefly describe some of the results in the context of 
abundance observations of EMP halo stars. This work
formed part of SWC's PhD thesis (completed in March 2007) and
a series of subsequent papers will describe the results of the study
in more detail.
\end{abstract}
\maketitle
%
%%%%%%%%%%%%%%%%%%%%%%%%%%%%%%%%%%%%%%%%%%%%
%% MAINMATTER
%%%%%%%%%%%%%%%%%%%%%%%%%%%%%%%%%%%%%%%%%%%%
%
%
%%%%%%%%
\section{Motivation}
\begin{figure}
\includegraphics[width=1\columnwidth]{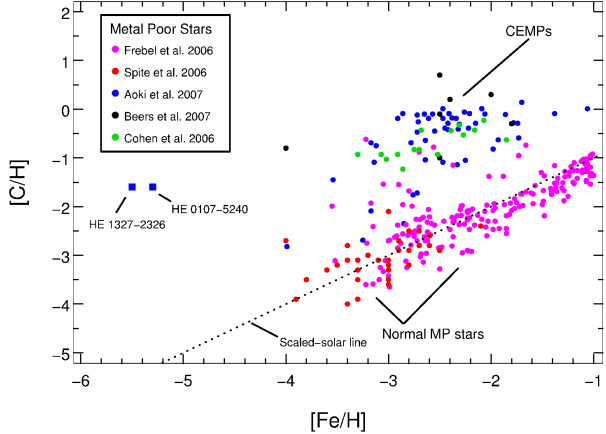}
\caption{Carbon abundance observations versus [Fe/H] for metal-poor stars
compiled from the literature. The dotted line marks $\textrm{[C/Fe]} = 0.0$.
The normal EMPs lie on this Galactic chemical evolution line, whilst
the CEMPs (which we define as having $\textrm{[C/Fe]} > +0.7$) lie
above this. The two most Fe-poor stars known, HE 0107-5240 ($\textrm{[Fe/H]} = -5.3$,
\cite{2004ApJ...603..708C}) and HE 1327-2326 ($\textrm{[Fe/H]} = -5.4$,
\cite{2006ApJ...639..897A}) are labelled.  
We note that this sample is biased towards CEMPs.
Observational data sets are from
\cite{2006ApJ...652.1585F}, \cite{2006AA...455..291S}, \cite{2007ApJ...655..492A},
\cite{2007AJ....133.1193B}, and \cite{2006AJ....132..137C}. \label{fig1}}
\end{figure}

%%%%%%%%
%
The discovery of extremely metal-poor stars (EMPs) in the Galactic Halo has 
renewed interest in the theoretical
modelling of Population III and low-metallicity stars. Most of these
stars show abundances that conform to a simple Galactic chemical evolution
line (see Figure \ref{fig1}). However a subset of the EMP stars have been 
observed to contain large amounts
of carbon. These C-rich EMPs (CEMPs) make up a large proportion of
the EMP population ($\sim 10 \rightarrow 20\%$; see eg. \cite{BC05}).
This population is also highlighted in Figure \ref{fig1}. Apart from
carbon the EMP stars also display variation in a range of other elements (see
\cite{BC05} for a review of the observations). A number of theories
have been proposed to explain the various patterns, ranging from pre-formation
pollution via Pop III supernovae (eg. \cite{1998ApJ...507L.135S},
\cite{2003ApJ...594L.123L}) to self-pollution through peculiar evolutionary
events (eg. \cite{2000ApJ...529L..25F}, \cite{2004AA...422..217W})
to binary mass transfer (eg. \cite{2004ApJ...611..476S}). 

In the current study we have undertaken a broad exploration of EMP
stellar evolution and nucleosynthesis in the low and intermediate
mass regime. Our study expands on the previous work in the field.
In particular it includes (for the first time) full evolutionary 
and nucleosynthesis calculations
from ZAMS to the end of the thermally-pulsing AGB phase (TPAGB), as
well as chemical yields for 74 nuclear species. With this homogeneous
set of models we hope to shed some light on whether or not 1D stellar
models can explain some of the EMP halo star observations.
%
%
%%%%%%%%
\section{Method}
%%%%%%%%
%
Our simulations were performed utilising two numerical codes -- a
stellar structure code and post-process nucleosynthesis code.

The stellar structure code used was the Monash/Mount Stromlo STAR
code ($\mbox{MONSTAR}$ see eg. \cite{1981ApJ...247..247W}, \cite{1996ApJ...473..383F}).
We do not describe the code in detail here but briefly note a few
key points. The code is largely a standard 1D code that utilises the
Henyey-matrix method (a modified Newton-Raphson method) for solving 
the stellar structure equations. Opacities
have been updated to those from \cite{1996ApJ...464..943I} (for mid-range temperatures)
and \cite{2005ApJ...623..585F} (for low temperatures). For the present
study the instantaneous convective mixing routine was replaced by
a time-dependent (diffusive) mixing routine (similar to that described
by \cite{2004AA...416.1023M}). This change was necessary due to
the violent evolutionary events that occur in models of EMP stars.
Convective boundaries were always defined by the Schwarzschild criterion
-- ie. no overshoot was applied. Thus the extension of convective
zones in all the models are conservative. A key problem with modelling
EMP stars is the unknown driver(s) of mass loss. The dominant theory
is that mass is lost from red giant envelopes through radiation
pressure acting on grains. Thus, in the EMP or $Z=0$ regime, mass
loss is thought to be negligible. The MONSTAR code uses empirical mass
loss formulae (the formula from \cite{1975MSRSL...8..369R} during
the RGB and that of \cite{1993ApJ...413..641V} during the AGB).
For this study we chose to retain the standard treatment, for the
following reasons. The first is that not much mass is lost during the RGB
 in these EMP models. This is because 
their RGB phases are much shorter (or non-existent), since He
is ignited much earlier than in stars of comparable mass at higher
metallicities. Indeed, the luminosity at the tip of the RGB can be
up to 1 order of magnitude lower in EMP models 
(see eg. \cite{1982AA...115L...1D}). Thus the RGB mass
loss is largely negligible, and it is the AGB mass loss that needs to
be handled properly. As described in the next Section these models
all experience some sort of polluting episode -- and always before
the AGB phase. This has the consequence that the surface of the AGB models
usually have metallicities approaching that of the LMC or even Solar
(as defined by Z rather than Fe -- they are still metal poor in terms
of Fe). Thus, since the stellar surfaces have (some of) the ingredients needed to form
grains, we argue that using the standard mass loss formula given by
\cite{1993ApJ...413..641V} is warranted. We also note that metallicity
is also indirectly taken into account by the mass loss formulae, since
they depend on bulk stellar properties (such as radius, luminosity,
pulsation period), which vary significantly with metallicity. 

The nucleosynthesis calculations were made with the Monash Stellar
Nucleosynthesis code ($\textrm{MONSN}$). MONSN is a post-process code.
As input it takes the key structural properties of each hydrostatic
model from the MONSTAR code (eg. density, temperature, convective
velocities). It solves a network of 506 nuclear reactions
involving 74 nuclear species (see eg. \cite{1993MNRAS.263..817C},
\cite{1996MmSAI..67..729L}, \cite{2004ApJ...615..934L} for more
details on this code).
Our grid of models covered the mass range: 
$\textrm{M} = 0.85,1.0,2.0,3.0 \textrm{M}_{\odot}$
and the metallicity range:
$\textrm{[Fe/H]} = -6.5,-5.45,-4.0,-3.0$, plus $Z=0$. 
%
%
%%%%%%%%
\section{The Dual Flash Events}
\begin{figure}
\includegraphics[width=1\columnwidth]{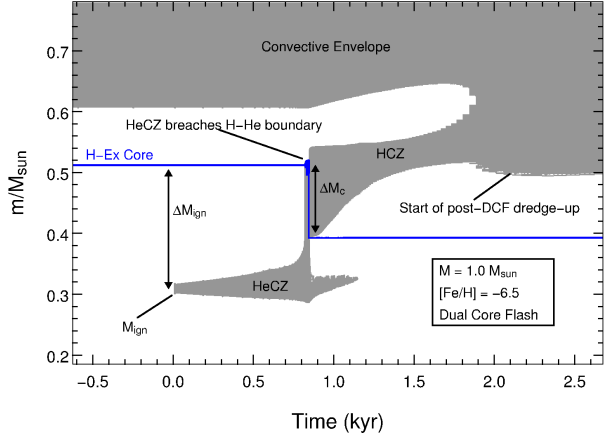}
\caption{An example of one of the Dual Core Flashes that occurred in some
of our low metallicity models. In this case it is the 1 M$_{\odot}$
model with {[}Fe/H]$=-6.5$. Convective regions are shown by grey
shading whilst the mass location of the edge of the H-exhausted core
is shown by the solid line (blue). The large inward movement of the
H-He boundary (H-Ex. core), due to the ingestion of protons into the
HeCZ, can be seen ($\Delta$M$_{c}$). The beginning of the post-DCF
dredge-up is indicated. This dredge up (mainly of CNO isotopes) increases
the surface metallicity to super-solar values (although [Fe/H] remains 
unchanged). \label{fig2}}
\end{figure}

%%%%%%%%
%
It has long been known that theoretical models of $Z=0$ stars (and EMP
stars) undergo violent evolutionary episodes not seen at higher metallicities.
This was first suggested by \cite{1982AA...115L...1D} and confirmed
by later calculations (see eg. \cite{1990ApJ...349..580F}, \cite{1990ApJ...351..245H},
\cite{1996ApJ...459..298C}). These episodes occur at different evolutionary
stages in stars of different mass and metallicity (see eg. \cite{2000ApJ...529L..25F}).
The most severe of these evolutionary events occurs during the core
He flash of low-mass stars (with $\textrm{[Fe/H]}<-2.5$,
see \cite{2000ApJ...529L..25F}). In this event the normal flash-driven
convective zone breaks out of the He-rich core. Thus H-rich material
is mixed down into regions of high temperature. Processed material
is also mixed upwards. The proton-rich material burns at a very high
rate, amounting to a secondary flash -- a H-flash. This flash reaches
luminosities comparable to the core He flash itself and occurs within
the same timeframe as the He flash. Thus we refer to the combination
of these events as a `Dual Core Flash' (DCF). We note that this event
has also been referred to as Helium Flash Induced Mixing (HEFM, \cite{2002AA...395...77S})
and Helium Flash-Driven Deep Mixing (He-FDDM, \cite{2004ApJ...611..476S}).
A similar event occurs in stellar models of higher mass and higher
(although still very low) metallicities. In these cases it is the
normal AGB shell He flash-driven convection zone that breaches the H-He
discontinuity. This occurs during the first few pulses of the TPAGB phase. Again a 
H-flash is induced during the evolution of
the He-flash, so we refer to this event as a `Dual \emph{Shell} Flash'
(DSF). Both the DCF and DSF events have consequences for the surface
composition of the star since, in both cases, the convective envelope
subsequently deepens and mixes up the (processed) material overlying
the H-burning shell. Presently there is reasonable consensus that
these events do indeed occur (at least in stellar models!), although
we note that not every study has found them to (eg. \cite{2002ApJ...570..329S}).
In Figure \ref{fig2} we display an example calculation of a DCF from
one of our models. 
%
%%%%%%%%
\section{Categorisation of Yields}
\begin{figure}
\includegraphics[width=1\columnwidth]{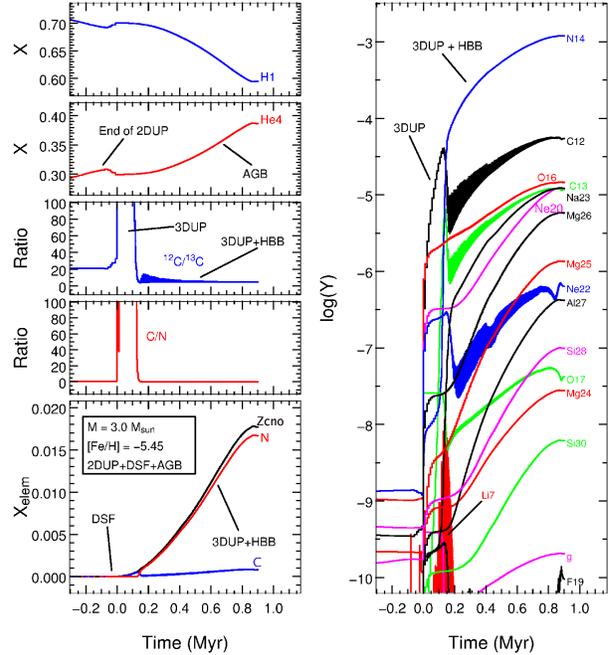}
\caption{The time evolution of the surface composition in the $\textrm{[Fe/H]}=-5.45$,
3 M$_{\odot}$ model,
for selected species. This model fits into our Group 3. 
The rich nucleosynthesis arising from 3DU and HBB is seen in the right-hand
panel. In particular the CN cycling product $^{14}$N dominates during
most of the AGB. The $^{12}$C$/^{13}$C and $\textrm{C/N}$
ratios quickly approach equilibrium values once the (strong) HBB starts.
This chemical signature by far dominates that of the DSF in this case.\label{fig3}}
\end{figure}

\begin{figure}
\includegraphics[width=1\columnwidth]{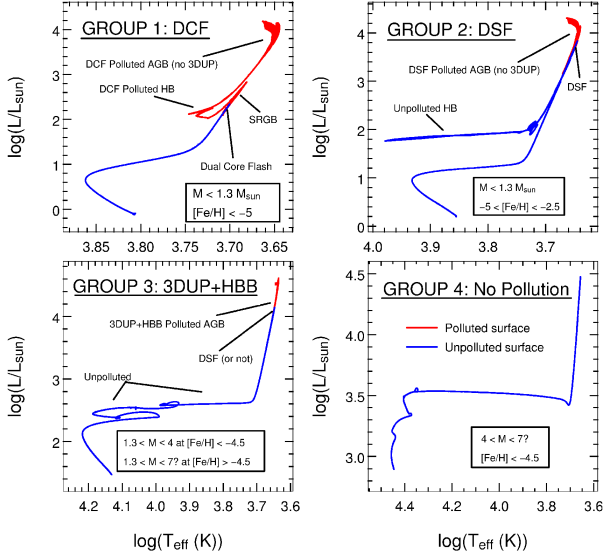}
\caption{Displayed in each
HR diagram is a representative example from our grid of models for
each pollution group.
Grey (red) lines indicate phases of the evolution in which the surface
is strongly polluted with CNO nuclides (from the DCF, DSF or 3DU events).
Black (blue) lines indicate that the surface still retains the initial
metal-poor composition. Evolutionary stages and pollution sources
are marked, as are the mass and metallicity ranges of each group.
Question marks indicate unknown upper boundaries (due to the limited
mass range of the current study). \label{fig4}}
\end{figure}

%%%%%%%%
%
In Figure \ref{fig3} we show the results from a nucleosynthesis calculation
for one of our models. In this model it is the Third Dredge-Up (3DU -- the periodic 
dredging up of He burning products into the convective envelope) and 
Hot Bottom Burning (HBB -- hydrogen burning at the bottom of the convective AGB 
envelope) that define
the yield of the star. Indeed, the chemical signature arising from
the DSF occurring at the start of the AGB is totally erased
by these normal AGB evolutionary episodes. However this is not always
the case. In Figure \ref{fig4} we summarise the pollution episodes
over the whole grid of models. We group the yields
into four categories, defined by the evolutionary events/phases that
dominate the chemical signature in the yields: 
Group 1 yields are dominated by the DCF events,
Group 2 are dominated by DSF events,
Group 3 are dominated by 3DU+HBB, whilst
Group 4 show no surface pollution during their hydrostatic evolution 
(but may explode as Type 1.5 supernovae).
Two key features visible in this figure are (1) members of the DCF group have
polluted surfaces during the horizontal branch phase (HB, the core He burning phase) 
and (2) both the DCF and DSF groups, which are of low
mass, have polluted surfaces during the AGB -- despite the lack of 3DU.
Thus our models predict a greater proportion of
C-rich stars at extremely low metallicity, as these events do not
occur at higher metallicities.
%
%
%%%%%%%%
\section{Comparisons with Observations}
\begin{figure}
\includegraphics[width=1\columnwidth]{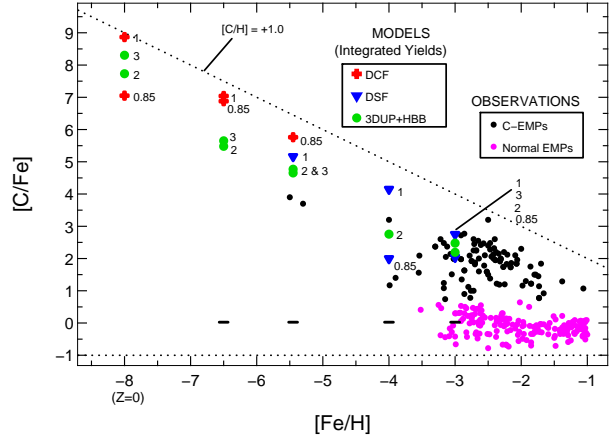}
\caption{Comparing the yields from all our models with observations of EMP
stars. See Figure \ref{fig1} for the observational data sources.
Here we have colour-coded the observations into [C/Fe]-rich (black dots) and
[C/Fe]-normal (grey/magenta dots), where [C/Fe]-rich is defined by $\textrm{[C/Fe]}>+0.7$.
The short horizontal lines indicate the starting composition of the
models (in this case they are all at $\textrm{[C/Fe]}\sim0$, except
for the $Z=0$ models). We have plotted the $Z=0$ model yields at $\textrm{[Fe/H]}=-8$
for comparison. The two most metal-poor stars can be seen at $\textrm{[Fe/H]}\sim-5.5$
(they are both C-rich). An upper envelope to the pollution of the
models -- and the observations -- is marked by the dotted line at
$\textrm{[C/H]}=+1.0$. The yields from our models are colour- and
shape-coded to highlight the different episodes that produced the
bulk of the pollution in each yield. Numbers beside each yield marker 
indicate initial stellar mass, in M$_\odot$. \label{fig5}}
\end{figure}

\begin{figure}
\includegraphics[width=1\columnwidth]{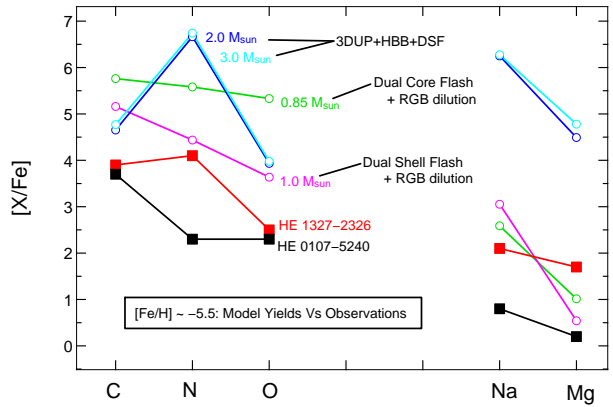}
\caption{Comparing the abundance pattern in our $\textrm{[Fe/H]}=-5.45$ yields
(open circles, initial stellar masses as indicated) with observations of stars 
with the same metallicity
(closed squares). The abundance determinations in these stars are
from \cite{2006ApJ...639..897A} (HE 1327-2326) and \cite{2004ApJ...603..708C}
(HE 0107-5240). \label{fig6}}
\end{figure}

%%%%%%%%
%
In Figure \ref{fig5} we compare the carbon yields from our entire
grid of models with the observed [C/Fe] abundances in EMP halo stars. It
can be seen that the yields from our models are all C-rich. Moreover,
the yields at $\textrm{[Fe/H]}=-3$, for which there are the most observations,
agree well with the observations. At $\textrm{[Fe/H]}=-4$ there is
a fair agreement with the observations, although there are less observations
to compare with. At $\textrm{[Fe/H]}=-5.45$ the yields show somewhat
more C than is observed, although there are only two stars observed
at this metallicity. An interesting feature of this diagram is that
the the model yields predict $\textrm{[C/Fe]}$ to continue increasing towards
lower metallicities. Furthermore, taking into account the evolutionary stage at
which the surface pollution is gained in the lower mass models ($\textrm{M}=0.85$
and 1.0 M$_{\odot}$) -- ie. the Dual Core Flash events rather
than the AGB -- the models also predict a higher proportion of C-rich
stars at lower and lower metallicities. This is due to the fact that
these stars already have polluted surfaces during the HB
stage -- which has a lifetime roughly 1 order of magnitude longer
than the AGB phase. 
The CEMPs also show interesting behaviour in other elemental abundances.
In Figure \ref{fig6} we compare the integrated yields from our $\mbox{[Fe/H]}=-5.45$
models with the observed abundance \emph{patterns} of the two most
metal-poor halo stars. Apart from N these two stars show similar 
abundance patterns. They both have roughly the same
C and O abundances and, although offset to lower abundances, the Na
and Mg abundances follow the same pattern. Nitrogen is 2 dex more
abundant in HE 1327-2326. The dominance of N over (enhanced) C and
O in HE 1327-2326 is reminiscent of the pattern that CNO cycling plus 3DU creates.
This pattern is also seen in the yields of the 2 M$_{\odot}$ and 3 M$_{\odot}$ models.
In these models it is 3DU+HBB that dominates the chemical signature
of the yields. It may be possible that HE 1327-2326 has been polluted
by an intermediate mass AGB star that had undergone 3DU and partial
(or lower temperature) HBB. Na and Mg are very much overproduced in
these models compared to the observations but, again, lower temperature
or incomplete HBB may account for this. HE 0107-5240 on the other
hand is better `matched' (it is still far from ideal) by the 1 M$_{\odot}$
yield. The yield of this model is dominated by the DSF event at the
start of the AGB. In this case N and Na are overproduced in
the model relative to the observations. 

More comparisons and analysis will be reported in future papers arising from this
thesis. 
%
%
%%%%%%%%
\section{Summary}
%%%%%%%%
%
In agreement with most previous studies we find that many of the 
EMP and $Z=0$ models experience
violent evolutionary episodes not seen at higher metallicities. We
refer to these events as `Dual Flashes' (DFs). 

The models predict an increased number of carbon-rich stars at the 
lowest metallicities. This is mainly due to the extra pollution provided by the
DF events. This concurs well with the observations, which show the proportion of 
CEMP stars in the Galactic halo to be higher at lower
metallicities. Although not discussed here we also note that we found 
the pollution arising from the DF events to be simultaneously C- and 
N-rich, as also observed in the CEMP stars. This contrasts with the pollution 
expected from 3DU at low mass, which would be (mainly) C-rich. Furthermore,
the models predict that the proportion of CEMP stars should \emph{continue}
to increase at lower metallicities, based on the results that the low mass 
EMP models already have polluted surfaces by the HB phase, 
and that there are more C-producing evolutionary episodes at these metallicities.
We also compared the chemical pollution arising from the models with 
the detailed abundance patterns available
for some of the most metal-poor CEMP stars, and found mixed results.

We note that all these calculations contain many uncertainties.
These include the unknown mass-loss rates, uncertain nuclear reaction
rates, and the treatment of convection. In the case of the DF
events a further uncertainty is the possibility that full fluid dynamics
calculations are really needed to model these violent episodes. 

The models and yields from this thesis will be described in more detail
in a series of future papers. For interested readers a full version of 
the thesis is available on SWC's webpage.
%
%
%%%%%%%%%%%%%%%%%%%%%%%%%%%%%%%%%%%%%%%%%%%%%%%%
%% BACKMATTER
%%%%%%%%%%%%%%%%%%%%%%%%%%%%%%%%%%%%%%%%%%%%%%%%
%
\begin{theacknowledgments}
Most of the calculations for this study utilised the Australian Partnership
for Advanced Computing's supercomputer, under Project Code \textit{g61}.
SWC thanks the original authors and maintainers of the stellar codes that have 
been used in this thesis -- Peter Wood, Rob Cannon, John Lattanzio, Maria Lugaro, Amanda Karakas,
Cheryl Frost, Don Faulkner and Bob Gingold. SWC was supported by a
Monash University Research Graduate School PhD scholarship
for 3.5 years.
\end{theacknowledgments}
%
%%%%%%%%%%%%%%%%%%%%%%%%%%%%%%%%%%%%%%%%%%%%%%%%
%% The bibliography can be prepared using the BibTeX program or
%% manually.
%%
%% The code below assumes that BibTeX is used.  If the bibliography is
%% produced without BibTeX comment out the following lines and see the
%% aipguide.pdf for further information.
%%
%% For your convenience a manually coded example is appended
%% after the \end{document}
%%%%%%%%%%%%%%%%%%%%%%%%%%%%%%%%%%%%%%%%%%%%%%%%

%%%%%%%%%%%%%%%%%%%%%%%%%%%%%%%%%%%%%%%%%%%%%%%%
%% You may have to change the BibTeX style below, depending on your
%% setup or preferences.
%%
%%
%% For The AIP proceedings layouts use either
%%%%%%%%%%%%%%%%%%%%%%%%%%%%%%%%%%%%%%%%%%%%

\bibliographystyle{aipproc}   % if natbib is available
%\bibliographystyle{aipprocl} % if natbib is missing

%%%%%%%%%%%%%%%%%%%%%%%%%%%%%%%%%%%%%%%%%%%
%% You probably want to use your own bibtex database here
%%%%%%%%%%%%%%%%%%%%%%%%%%%%%%%%%%%%%%%%%%%
\bibliography{fs3}

%%%%%%%%%%%%%%%%%%%%%%%%%%%%%%%%%%%%%%%%%%%
%% Just a reminder that you may have to run bibtex
%% All of it up to \end{document} can be removed
%% if you don't like the warning.
%%%%%%%%%%%%%%%%%%%%%%%%%%%%%%%%%%%%%%%%%%%
\IfFileExists{\jobname.bbl}{}
 {\typeout{}
  \typeout{******************************************}
  \typeout{** Please run "bibtex \jobname" to obtain}
  \typeout{** the bibliography and then re-run LaTeX}
  \typeout{** twice to fix the references!}
  \typeout{******************************************}
  \typeout{}
 }

\end{document}